\documentclass[cits]{PoS}
\usepackage{amsmath, amssymb, bm}
\usepackage{subfigure}

\newcommand{\preprintline}{\newline
\vskip -5.2cm
\rightline{\parbox{5cm}{\large\rm CERN-PH-TH-2010-260\\ HU-EP-10/68\\ SFB/CPP-10-95}}
\vspace{4.2cm}}

\title{Comparison of the mass preconditioned HMC and the DD-HMC algorithm for two-flavour QCD \preprintline}

\ShortTitle{Comparison of mass preconditioned HMC and DD-HMC}

\author{\speaker{Marina Marinkovic} \\
        Humboldt Universit\"at zu Berlin, Institut f\"ur Physik, 
	Newtonstr. 15, 12489 Berlin, Germany\\ 
        E-mail: \email{mmarina@physik.hu-berlin.de}}

\author{Stefan Schaefer\\
        Humboldt Universit\"at zu Berlin, Institut f\"ur Physik, 
	Newtonstr. 15, 12489 Berlin, Germany\\ 
 CERN, Physics Department, 1211 Geneva 23, Switzerland \\ 
        E-mail: \email{stefan.schaefer@cern.ch}}

\abstract{Mass preconditioned HMC and DD-HMC are among the most popular algorithms\
to simulate Wilson fermions. We present a comparison of the performance\
of the two algorithms for realistic quark masses and lattice sizes. In particular, 
we use the locally deflated solver of the DD-HMC environment also for the mass 
preconditioned simulations.}

\FullConference{The XXVIII International Symposium on Lattice Field Theory, Lattice2010\\
		June 14-19, 2010\\
		Villasimius, Italy}

\begin{document}
%%%%%%%%%%%%%%%%%%%%%%%%%%%%%%%%%%%%%%%%%%%%%%%%%%%%%%%%%%%%%%%%%%%%%%%%%%%%%%%%%%%%%%%%%
\section{Introduction}
%%%%%%%%%%%%%%%%%%%%%%%%%%%%%%%%%%%%%%%%%%%%%%%%%%%%%%%%%%%%%%%%%%%%%%%%%%%%%%%%%%%%%%%%%
In order to accelerate numerical simulations in lattice QCD, different 
preconditioning techniques are being used. We will try to make a comparison between
two popular  ways 
of preconditioning the Hybrid Monte Carlo (HMC) for improved Wilson fermions: domain 
decomposition introduced by L\"uscher in the DD-HMC algorithm\cite{Luscher:2005rx} 
and  Hasenbusch's mass preconditioning 
(MP)\cite{Hasenbusch:2001ne}. In both approaches, the quark determinant is factorized into 
a part which is dominated by the infra-red and another which is largely ultra-violet.
This leads to the reduction of the quark force magnitude 
in the molecular dynamics equations. Therefore, the associated integration step sizes can 
be set to larger values, which gives an acceleration of the  algorithm.

Comparisons between algorithms are difficult. In a modern  lattice QCD
computation, many parameters have to be set. Since their number can go into
the dozens---their optimal values depending on each other---it is virtually
impossible to find the minimum, at which each algorithm performs best, and
then make a true comparison. In particular, the performance is determined by the 
auto-correlation times of the observables of interest, whose measurements require 
runs with high statistics. Furthermore, the  optimal values of the parameters and 
the relative performance of the algorithms might also
depend on the implementation and the computer the simulation is run on.

At least from the point of view of implementation, we tried to have the
comparison on virtually equal grounds: for the DD-HMC, we used L\"uscher's
publicly available code\footnote{http://cern.ch/luscher/dd-hmc} and from it,
we developed an implementation of the mass preconditioned HMC, reusing as many
building blocks  as possible.  In particular the locally deflated solver
introduced in the DD-HMC framework\cite{Luscher:2007se} turns out to greatly
speed up the simulations in both algorithmic setups.  

The block decomposition, on which the DD-HMC is based, allows for a decoupling of
the blocks for a large part of the forces in such a simulation. The links on
the boundary are not updated during the trajectory, which  results in reduced
communication, and is therefore suitable for clusters with fast nodes and a
relatively slow connection.  However, this also means that only a fraction $R$
of the links is ``active''. The naive expectation that the auto-correlation
times grow proportional to $R^{-1}$ is confirmed in pure gauge
theory\cite{Schaefer:2010hu}, however, with dynamical fermions the cost is not
reduced accordingly. There is a competition between the need of the computer,
which are small blocks using many processors of massively parallel machines, and
the need of the physics, which asks for blocks of a physical size of at least
$0.5$fm to provide an efficient preconditioning. The size of the blocks will
determine the relative size of the block force and the global remainder.

In mass preconditioning, the fermion matrix is preconditioned (basically) 
by one with a fermion of a larger mass. The value of its mass plays the role 
of the size of the block in domain decomposition, but
it is continuous  and therefore can be tuned. Also more than one preconditioning
fermion can be easily used.

Both algorithms have been compared in \cite{Jansen:2005yp}, however, new to the present
study is the use of the deflated solver in both algorithms.
The DD-HMC is well documented in the literature, so we will only describe our setup for 
the  MP-HMC algorithm in Section \ref{s:alg} and \ref{s:parm}, and  continue to our test 
in Sect. \ref{s:sol}, \ref{s:test1} and \ref{s:test2}.
%%%%%%%%%%%%%%%%%%%%%%%%%%%%%%%%%%%%%%%%%%%%%%%%%%%%%%%%%%%%%%%%%%%%%%%%%%%%%%%%%%%%%%%%%%
\section{Action and algorithms}
\label{s:alg}
%%%%%%%%%%%%%%%%%%%%%%%%%%%%%%%%%%%%%%%%%%%%%%%%%%%%%%%%%%%%%%%%%%%%%%%%%%%%%%%%%%%%%%%%%%
We are simulating $N_f=2$ degenerate flavours of non-perturbatively O($a$) improved Wilson 
quarks, using the Wilson gauge action. The Dirac operator in this formulation is given by
\begin{equation}
D(m) =D_W+c_{sw} \sum\nolimits_{\mu,\nu}^{~} \tfrac{i}{4} \sigma_{\mu \nu} \hat{F}_{\mu \nu}+m \ ,
\end{equation}
 where $D_W$ represents the
unimproved Wilson Dirac operator without mass term, $c_{sw}$ is the improvement coefficient and 
$m$ the bare quark mass. 

In our implementation of MP-HMC, we use mass preconditioning for the Schur complement of the 
symmetric even-odd preconditioning.
Starting from the standard decomposition,
\begin{equation}
\det Q = \det(\gamma_5 D)= \det Q_{ee}~\det {Q_{oo}}~\det {{Q}_S} \quad
\text{with}\quad Q_S = 1-Q_{ee}^{-1} Q_{eo} Q_{oo}^{-1} Q_{oe} \ ,
\end{equation}
we write
\begin{equation}
\det Q =  \det Q_{ee}~\det {Q_{oo}}~\det \big [ W(\Delta m) \big ]~\det \big [ W^{-1}(\Delta m)Q_S \big ]
\end{equation}
where $W(\Delta m)=Q_S + \Delta m$ with $\Delta m>0$. This leads to the effective action
\begin{equation}
S_\mathrm{eff} = -2(\log \det Q_{ee} + \log \det Q_{oo}) + |W^{-1}(\Delta m)~ \Phi_1|^2 +|(1+\frac{\Delta m}{Q_S}) \Phi_2|^2 \ .
\end{equation}
 Using again a Schur complement approach,  the inverse of the 
preconditioned operator $Q_S$ can be constructed from the  inverse of the full Hermitian Dirac operator $Q$
\begin{equation}
Q_S^{-1}=P_e {Q}^{-1} Q_{ee} P_e.
\end{equation}
In the following, the forces associated with pseudo-fermion field $\Phi_1$ are denoted by $F_1$, whereas $F_2$ are the forces from $\Phi_2$. 

In the DD-HMC, the quark determinant is written as the product of the determinants of the 
Dirac operator restricted to the blocks and a factor which accounts for the remaining
contributions to the fermion determinant. The latter factor couples the gauge fields on the different blocks.
The block forces are referred to as $F_1$ and $F_2$ is the block interaction force. 
For details of this setup see \cite{Luscher:2005rx}.

In both setups,  UV/IR separation due to the preconditioning opens the possibility to integrate $F_2$ 
on a larger time scale than $F_1$.

%%%%%%%%%%%%%%%%%%%%%%%%%%%%%%%%%%%%%%%%%%%%%%%%%%%%%%%%%%%%%%%%%%%%%%%%%%%%%%%%%%%%%%%%%%
\section{Simulation parameters}
\label{s:parm}
%%%%%%%%%%%%%%%%%%%%%%%%%%%%%%%%%%%%%%%%%%%%%%%%%%%%%%%%%%%%%%%%%%%%%%%%%%%%%%%%%%%%%%%%%%
We have performed runs on a $48\times24^3$ lattice at
$\beta=5.3$ and  $c_{sw}=1.90952$, corresponding to the lattice spacing in
physical units of  $a\approx 0.071$fm from $r_0=0.5$fm \cite{Leder:Latt2010}.
The hopping parameter $\kappa_\mathrm{sea}=0.13625$ corresponds to a pion mass of 
$m_{\pi}\approx 420 \mathrm{MeV}$ and $m_{\pi}L \approx 3.6$.  In all our runs, the trajectory
length is set to $\tau = 0.5$, for which we take the normalization of 
Ref.~\cite{Luscher:2005rx}.
These parameters were also used in DD-HMC simulations without deflation in 
Ref.~\cite{DelDebbio:2006cn}.

In DD-HMC,  the locally deflated Schwarz-preconditioned generalized
conjugate residual (GCR) solver described in \cite{Luscher:2007se} is employed for the inversions in $F_2$. The 
less expensive inversions on the blocks, needed in the computation of $F_1$, are done with 
the BiCGstab algorithm.
In the multiple time scale integration we use $N_1$ steps in the fermion force $F_1$ for
each of the $N_2$ steps per trajectory of $F_2$ and analogously $N_0$ steps of the gauge force per step in $F_1$.
We choose $N_2$, $N_1$ and $N_0$ to be 18, 5 and 6, respectively, for a block size of $6^2\times12^2$.

Without much tuning, for the MP case we have taken the same step size at the outer force 
and the rest of the parameters is chosen according to the ratio of the forces magnitude, 
i.e. $\|F_2\|:\|F_1\|:\|F_0\| = 1 : 9 : 36$. To be on the safe side, we have chosen 
$N_2$, $N_1$, $N_0$ = 18, 10, 10. In our version of MP-HMC,  the Schwarz-preconditioned GCR solver is employed 
for the computation of both $F_1$ and $F_2$. The demanding 
inversions with the low quark mass in the $F_2$ computation are done with the deflated version 
of the solver whereas in the  $F_1$ computations the deflation was off. Here, the preconditioning 
parameter is the positive mass difference added to the symmetrically preconditioned Dirac
operator which we set to $\Delta m~\approx~0.09$. More tuning could probably
lead to a better performance than the one discussed below.

In both setups, the standard leap frog integration is implemented and for the prediction of 
the solution in all inversions, the chronological inversion method of Brower et al. 
\cite{Brower:1995vx} is used.
%%%%%%%%%%%%%%%%%%%%%%%%%%%%%%%%%%%%%%%%%%%%%%%%%%%%%%%%%%%%%%%%%%%%%%%%%%%%%%%%%%%%%%%%%%
\section{Solver performance}
\label{s:sol}
%%%%%%%%%%%%%%%%%%%%%%%%%%%%%%%%%%%%%%%%%%%%%%%%%%%%%%%%%%%%%%%%%%%%%%%%%%%%%%%%%%%%%%%%%%

The Schwarz-preconditioned GCR solver used for all the inversions in MP-HMC is taken over
from the DD-HMC environment. The results of intensive tests of this solver implementation within
DD-HMC can be found in \cite{Luscher:2003qa,Luscher:2007es}. As expected, we find the improved
performance of the deflated solver also in the MP-HMC, gaining roughly  a
factor of three in the time needed for the computation of $F_2$ on our lattices compared to the case without 
the application of deflation, details can be found in Table \ref{tab:defl}.
%\begin{center}
\begin{table}
\begin{small}
\begin{center}
\begin{tabular}[h]{| c| c| c| }
		\hline 
		 $~ $ &\ \   without defl.\ \  & \ \ with defl.\ \   \\ \hline \hline
		 $\ \langle N_\mathrm{GCR} (F_2)\  \rangle$ & 107 & 23  \\  \hline
		  \ comp. time($F_2$)\ & 780.42s  & 264.98 s\\
		\hline
		\end{tabular}
		\end{center}
	    \end{small}
\caption{Average number of GCR iterations per trajectory and the total execution time for the $F_2$ computation 
in MP-HMC with the deflated solver and without the application of deflation. The size of the deflation blocks is 
$6^2\times4^2$. Note that a reduction in average time by a factor $\sim 3$ is achieved, taking into 
account the time needed for the construction of the deflation subspace.}
\label{tab:defl}
\end{table}
%\end{center}
%\vspace*{-20.0mm}
\begin{figure}[hd]
%\vspace*{-5.0mm}
\begin{center}
\includegraphics[scale=0.30,angle=270]{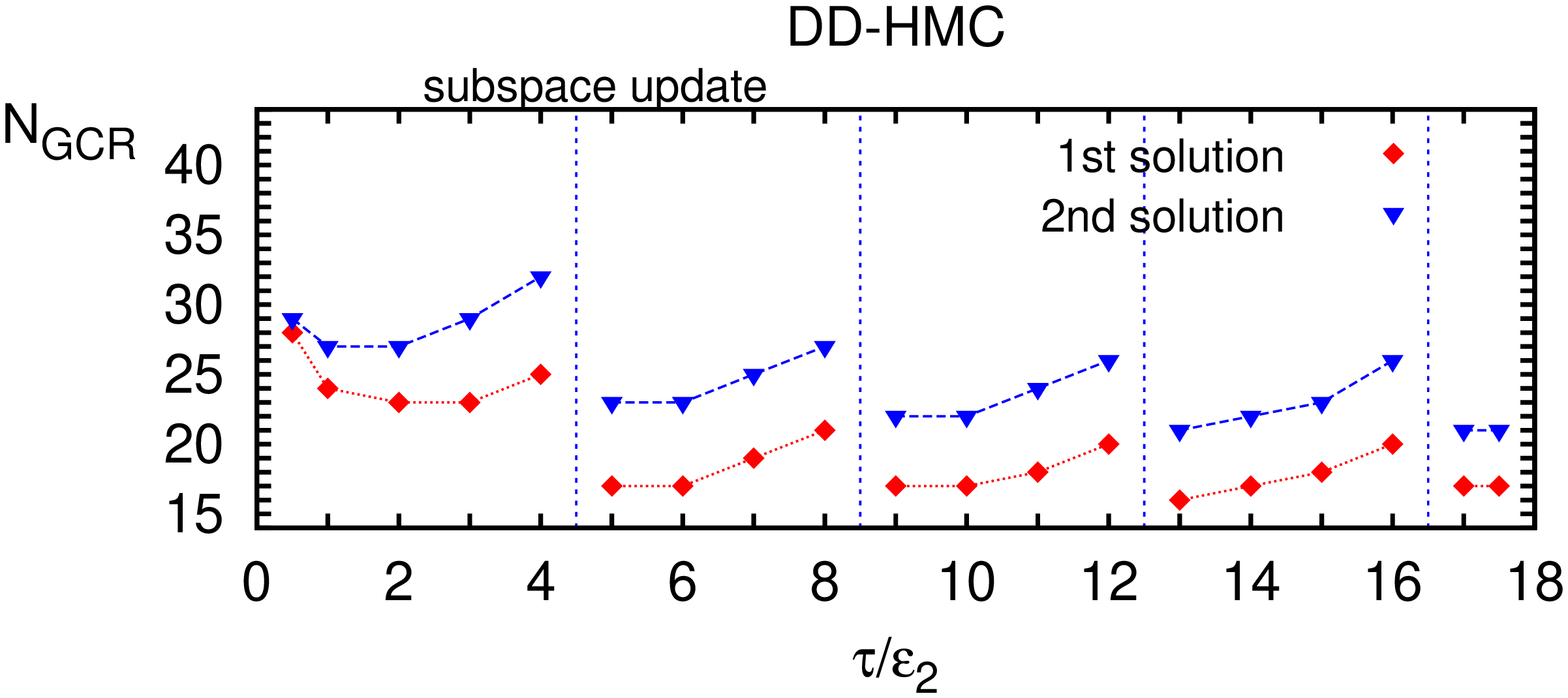}
\includegraphics[scale=0.30,angle=270]{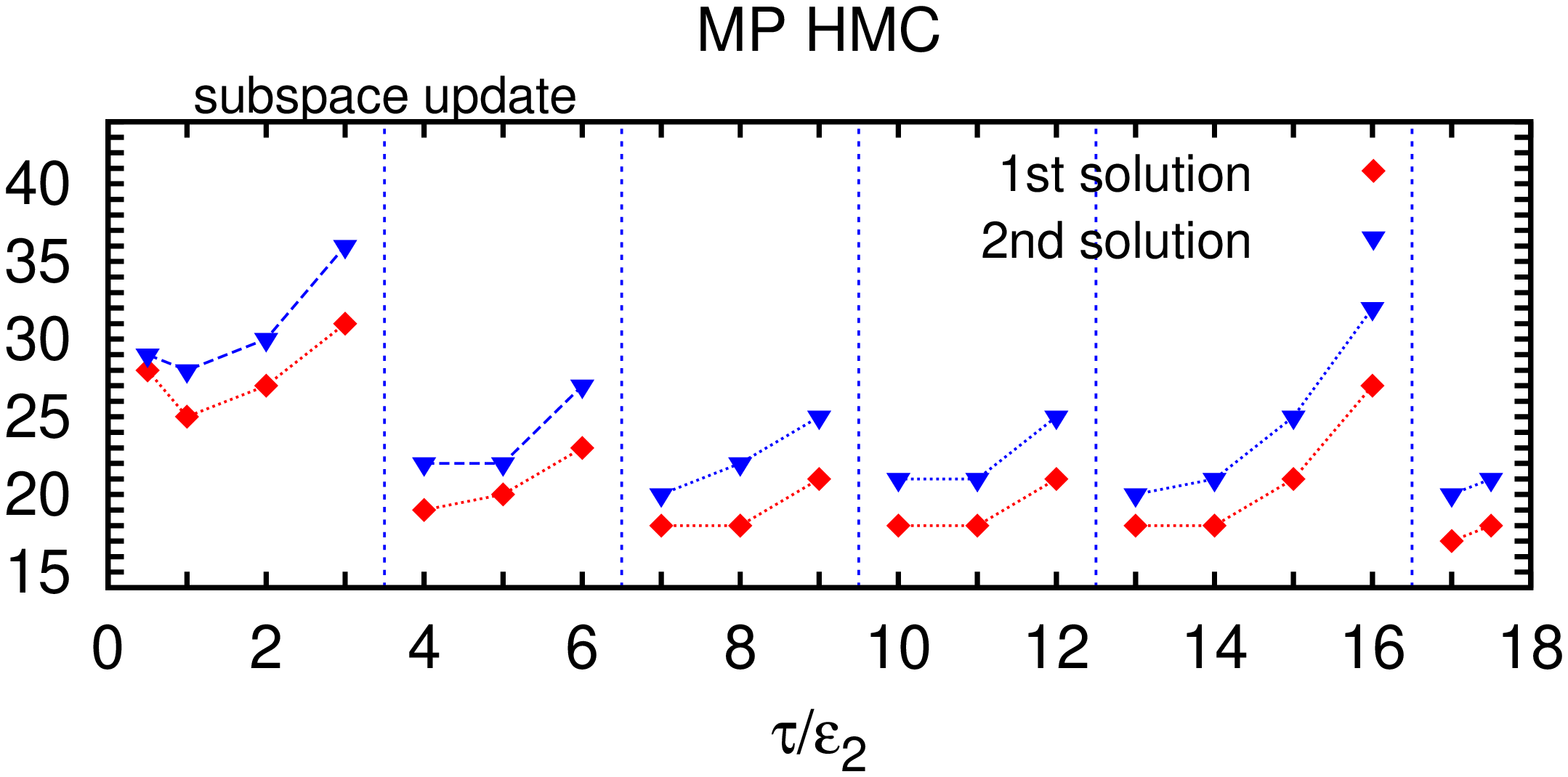}
\end{center}
%\vspace*{-5mm}
\caption{History of the iteration numbers $N_\mathrm{GCR}$ of the deflated 
Schwarz-preconditioned GCR solver along a molecular-dynamics trajectory, using the DD-HMC 
(left side) and MP-HMC (right side) algorithm, plotted against the molecular-dynamics 
time $\tau$ given in units of the integration step size ${\varepsilon}_2=\frac{\tau}{N2}$. 
The lattice size in both cases is $48\times24^3$ and $\kappa_\mathrm{sea}=0.13625$. The
vertical lines indicate the refreshment of the deflation subspace.}
\label{fig:ngcr}
\end{figure}

For the update of the deflation subspace the same criteria as in the DD-HMC 
setup were applied\cite{Luscher:2007es}, however, since in  MP-HMC all links are active,
the deflation subspace has to be updated more often than in the DD-HMC case to
satisfy the same update criteria, see Fig.~\ref{fig:ngcr}. 
In principle, we could have optimized the update criterion of the deflation space even
further to match the particular conditions in the MP-HMC given by the cost of the
$F_2$ force computation, but we have already achieved a very significant gain and do not
expect any further improvements to dramatically change the situation.

%%%%%%%%%%%%%%%%%%%%%%%%%%%%%%%%%%%%%%%%%%%%%%%%%%%%%%%%%%%%%%%%%%%%%%%%%%%%%%%%%%%%%%%%%%
\section{Quark forces and stability}
\label{s:test1}
%%%%%%%%%%%%%%%%%%%%%%%%%%%%%%%%%%%%%%%%%%%%%%%%%%%%%%%%%%%%%%%%%%%%%%%%%%%%%%%%%%%%%%%%%%

At small quark masses,  instabilities in the numerical integration of the molecular 
dynamics equations may occur, which manifest themselves in violent fluctuations in 
the energy violation $\Delta H$ of the molecular dynamics evolution.
According to the tests of the DD-HMC algorithm performed so far, 
severe integration instabilities were rare \cite{DelDebbio:2006cn}. This has to be demonstrated
also for the MP-HMC and for both algorithms when going to smaller quark masses.

\begin{figure}[*hb]
\subfigure{}{
\includegraphics[scale=0.29,angle=270]{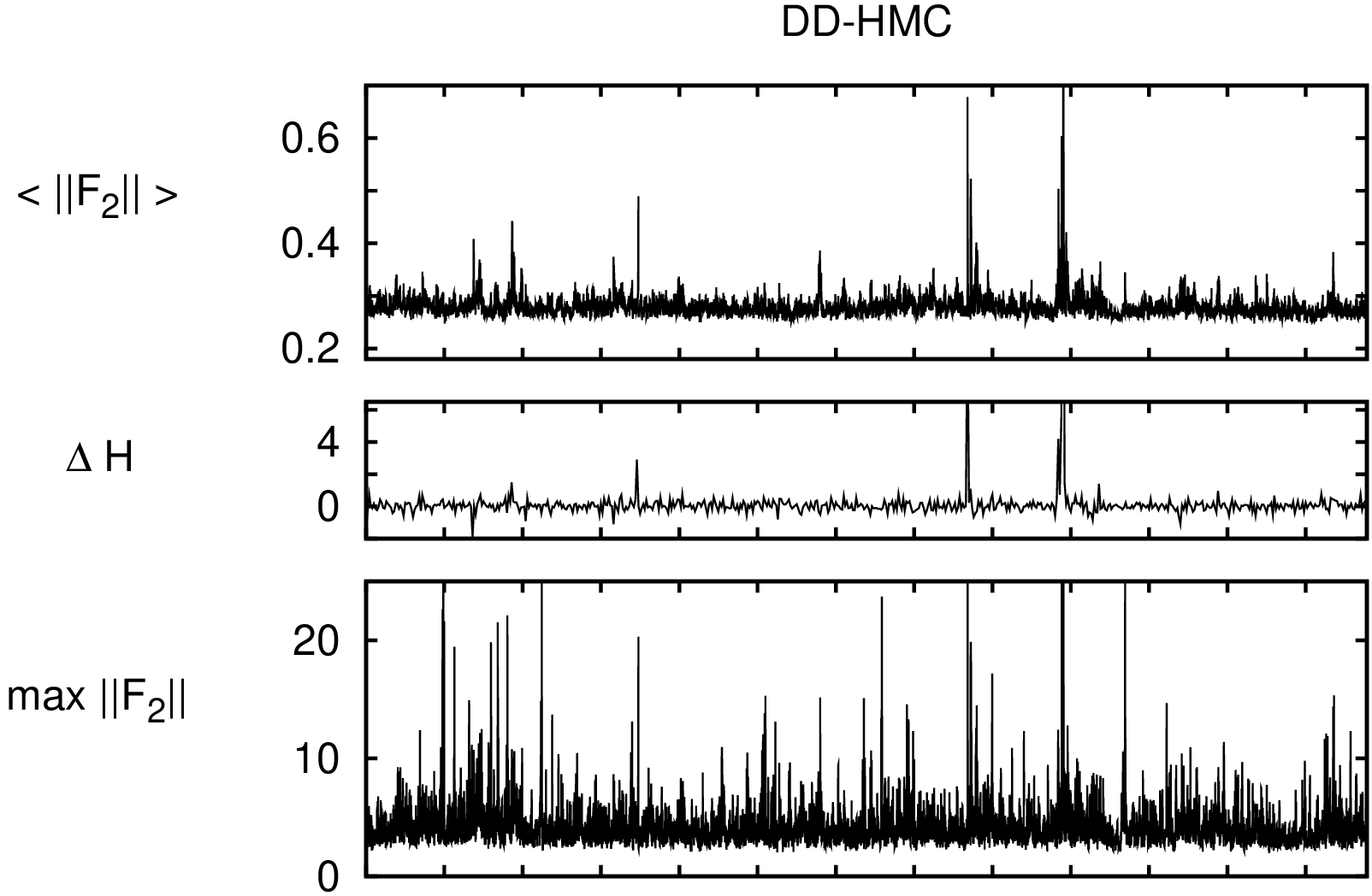}
%\label{fig:1}
}
\subfigure{}{
\includegraphics[scale=0.29,angle=270]{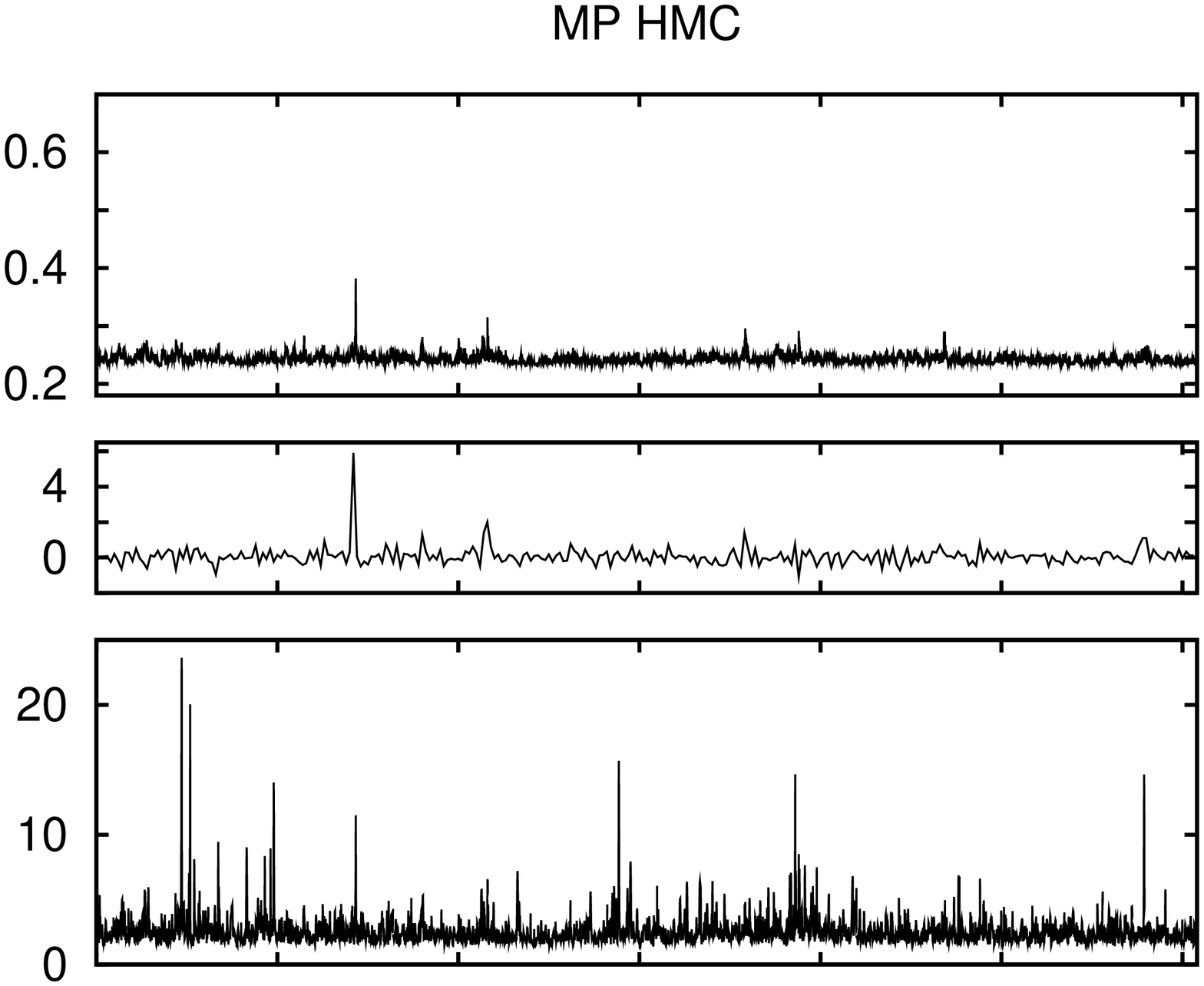}
%\label{fig:2}
}

\subfigure{}{
\includegraphics[scale=0.29,angle=270]{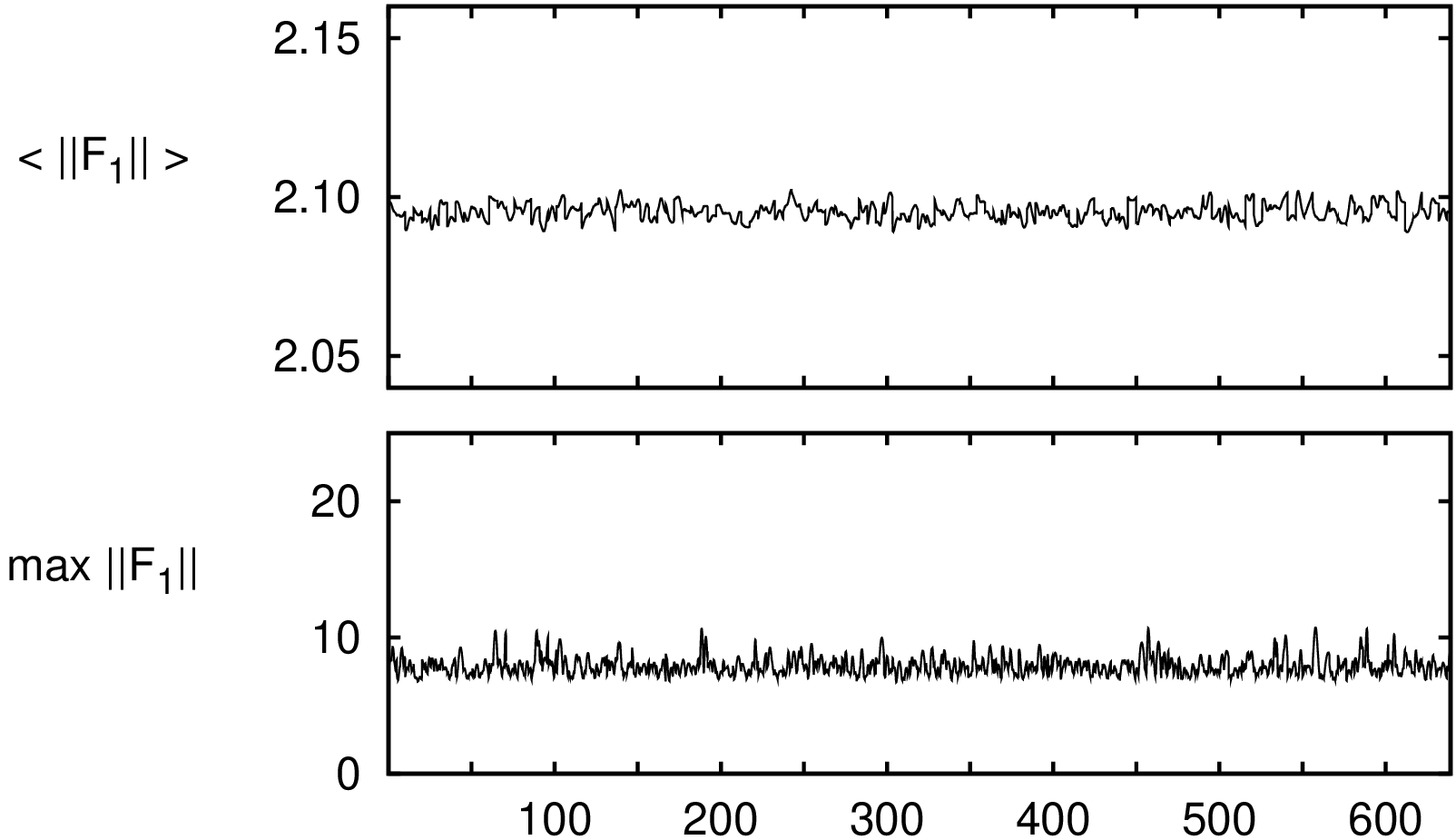}
%\label{fig:3}
}
\subfigure{}{
\includegraphics[scale=0.29,angle=270]{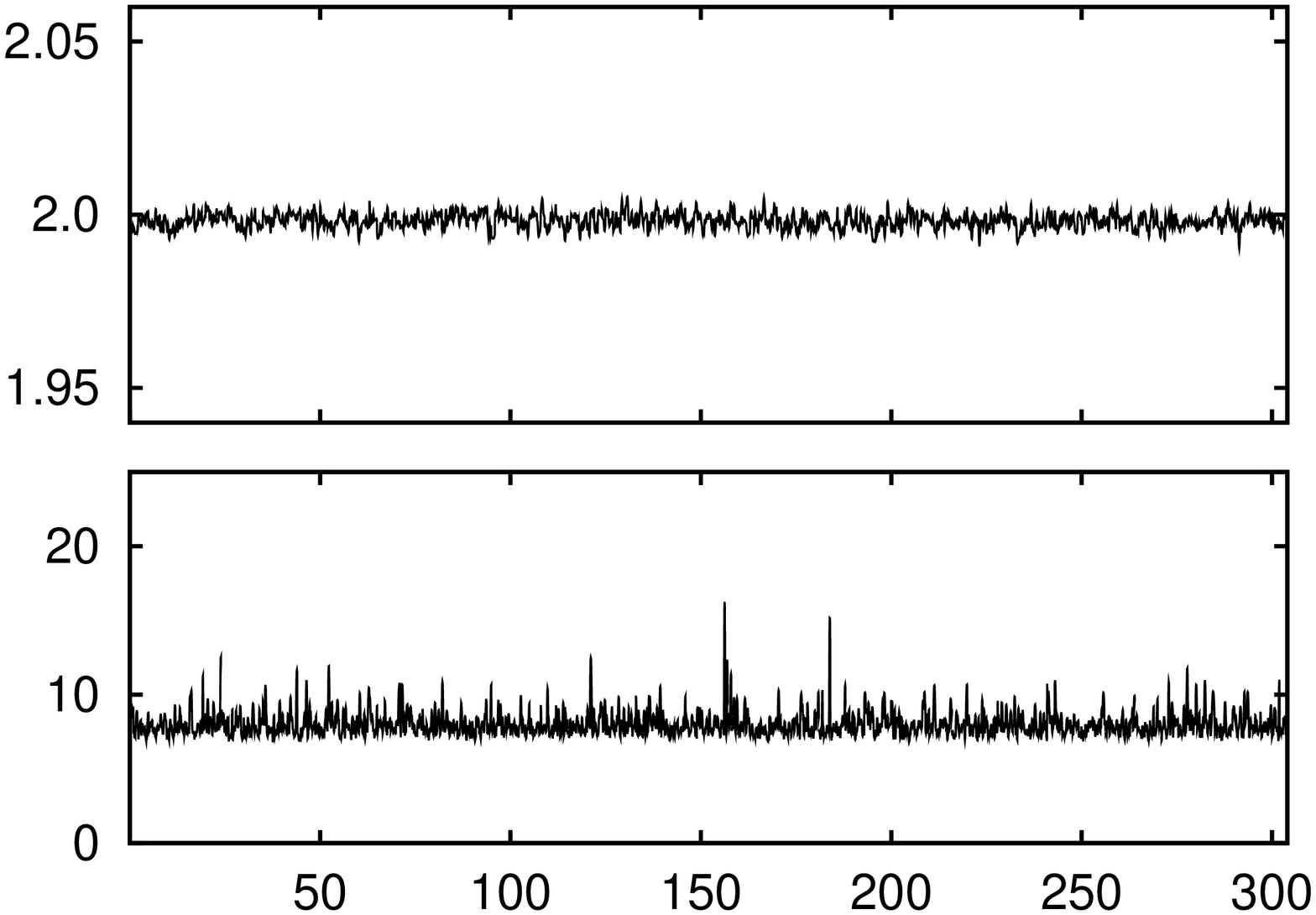}
%\label{fig:4}
}
\caption{Histories of the energy violation $\Delta H$, as well as maximum and average forces $F_2$ and $F_1$, for each
force update, plotted as a function of the trajectory number. Values corresponding to the DD-HMC 
algorithm are shown to the left and the integration stepsizes for the two forces relate as 
$\Delta t_2 : \Delta t_1 = 1:6$. The values for MP-HMC are shown in the two right panels and 
the corresponding ratio of the integration steps is $\Delta t_1 : \Delta t_2 = 1:10$. The lattice 
size is $48\times24^3$ and $\kappa_\mathrm{sea}=0.13625$.}
\label{fig:forces}
\end{figure}

Typically, the force $F_2$ is the source of these instabilities, and in Fig.~\ref{fig:forces} 
we show  Monte Carlo time histories for $F_1$ and $F_2$ showing the  maximal and average forces, 
together with the corresponding $\Delta H$ throughout roughly $600$ trajectories for DD-HMC and 
$300$ in the MP-HMC case.
\noindent 
One can see that the average force in the case of MP-HMC fluctuates a lot less than in 
DD-HMC, which is reflected in smaller magnitude and fluctuations in  $\Delta H$ for the mass
preconditioning. 

%%%%%%%%%%%%%%%%%%%%%%%%%%%%%%%%%%%%%%%%%%%%%%%%%%%%%%%%%%%%%%%%%%%%%%%%%%%%%%%%%%%%%%%%%%
\section{Efficiency of the algorithms}
\label{s:test2}
%%%%%%%%%%%%%%%%%%%%%%%%%%%%%%%%%%%%%%%%%%%%%%%%%%%%%%%%%%%%%%%%%%%%%%%%%%%%%%%%%%%%%%%%%%
The question of performance of the two algorithms must be addressed empirically and although
a final answer would require to include in the comparison a series of lattice sizes and quark
masses, our study could give us a first insight into how the two approaches relate in 
computational cost. This includes not only the cost of performing a single trajectory,
but also the related auto-correlation times, because what matters is the cost of achieving
a certain statistical error.
The presented computations are performed on the SGI Altix which is built of Intel Xeon Gainestown 
X5570 CPUs with InfiniBand connections at HLRN supercomputing system at ZIB in Berlin and RRZN in Hannover, and 
the relative timings can certainly be different on other architectures.

In Table~\ref{tab:DDvsMP} we show the average plaquette value in both runs, integrated 
autocorrelation time $\tau_\mathrm{int}$ and the acceptance rate for the two cases.
\begin{table}
\begin{small}
\begin{center}
	\begin{tabular}[h]{|c|c|c|c|c|c|}
	\hline
	$~ $ &  wall clock / R &$U_P$& $\tau_\mathrm{int}(U_P) \times R$ & $N_{traj}$&acc. rate \\ \hline \hline
	DD-HMC & 2010s &1.65106(10)  & 10(5)  & 840 &90(1)\% \\ 	\hline
	MP-HMC & 1530s &1.65127(10) & 10(4) & 432 &85(2)\% \\ 
	\hline
	\end{tabular}
\end{center}
\caption{Comparison of the DD-HMC algorithm with the MP-HMC. Both 
simulations are done for the improved Wilson theory with two degenerate fermion flavours. The 
lattice size is  $48 \times 24 ^ 3$, lattice spacing $a\approx 0.07$ fm and the pion mass 
$m_\pi \approx 400$ MeV. The block size in DD-HMC is $6^2\times12^2$, while in the MP case, the 
difference in mass $\Delta m \sim 0.09$. Here $R$ represents the fraction 
of active links in the algorithm, $R=0.36$ for DD-HMC and $R=1$ for MP-HMC. The two algorithm 
demonstrate comparable performance.
}
\label{tab:DDvsMP}
\end{small}
\end{table}
\noindent
As we have already mentioned, the fraction of active links in the DD-HMC for the chosen 
parameters is $R=36\%$, increasing it would mean to run with fewer than 32 MPI processes.
 Note, however, that this is partially due to the relatively small $L/a=24$.
It has been shown for the pure gauge theory that the autocorrelation time scales with the inverse 
of this fraction\cite{Schaefer:2010hu} as long as the blocks are of a reasonable size.
In order to be able to compare the two algorithms, we have multiplied 
the current result for the integrated autocorrelation time with 
$R$, and scaled the execution time accordingly. Since the available statistics 
is not enough for the reliable estimation of the errors in the autocorrelation times, the 
values for the $\tau_\mathrm{int}$ and its error should be taken only as first estimates. Including 
the acceptance into consideration, we can conclude from the performed study that in both 
approaches roughly the same CPU time is needed for the same error in the measured observable $U_P$.
%%%%%%%%%%%%%%%%%%%%%%%%%%%%%%%%%%%%%%%%%%%%%%%%%%%%%%%%%%%%%%%%%%%%%%%%%%%%%%%%%%%%%%%%%
\section{Summary }
%%%%%%%%%%%%%%%%%%%%%%%%%%%%%%%%%%%%%%%%%%%%%%%%%%%%%%%%%%%%%%%%%%%%%%%%%%%%%%%%%%%%%%%%%

In this contribution, we have compared DD-HMC, one of the most efficient algorithms for
dynamical QCD simulations,  with our implementation of a mass preconditioned HMC, 
including for the first time a locally deflated solver, which brings
a significant speedup. We have confirmed that relatively
large step size for small quark mass is achievable also with MP-HMC. Looking at
the series of average and maximal forces of $F_2$ in both cases, it is
indicative that the energy violations visible as spikes in $\Delta H$ are
caused by the irregularities in the forces. This is easier to control with
a continuous parameter, such as the preconditioning mass in mass preconditioning case,
than with the  HMC block size in
DD-HMC which can only take few values in practical applications.
 The two algorithms have demonstrated comparable performance, which is
in large owed to the usage of the same efficient deflated solver in both cases. 

A future task is to extend this study to larger lattices with  even lower quark
masses. The MP-HMC also leaves room for improvement. One could study the use of
three or more pseudo-fermion fields and also tune the preconditioning masses more
carefully than we did in the present results.  In particular for smaller
preconditioning masses, the use of the deflated solver also for $F_1$ might be advisable.

Besides being able to use larger numbers of CPUs, the approach with mass preconditioning 
allows for a much easier extension of the program package, such as introducing Schr\"odinger 
functional boundary conditions, as well as adding additional heavy and non-degenerate flavours. 

\section{Acknowledgements}

We thank M. Hasenbusch, H. Simma, R. Sommer, F. Virotta and U. Wolff for
useful and stimulating discussions and M.~L\"uscher for making his DD-HMC code
publicly available. This work is supported by the German Science Foundation (DFG) 
under the grants GRK1504 "Mass, spectrum, symmetry" and SFB/TR9-03. Simulations
were performed at HLRN  in Berlin and Hannover.

\end{document}